\documentclass[aps,prl,reprint,superscriptaddress,footinbib,floatfix]{revtex4-2}

\usepackage{natbib}
\usepackage{amsmath,amssymb,bm}
\usepackage{booktabs}
\usepackage{graphicx}
\usepackage{microtype}
\usepackage{xcolor}
\usepackage{txfonts}
\usepackage[breaklinks,colorlinks,linkcolor=blue,citecolor=blue,urlcolor=blue]{hyperref}

\newcommand{\ii}{\mathrm{i}}

\newcommand{\fAthena}{0.2194}
\newcommand{\omegaTwoFive}{0.3496}
\newcommand{\omegaThree}{0.3071}

\newcommand{\figOneDistance}{2.5}
\newcommand{\figOneWindowStart}{20}
\newcommand{\figOneWindowEnd}{140}
\newcommand{\figOneDriveFrequency}{0.1113}
\newcommand{\figOneOrbitalAngularFrequency}{0.3496}
\newcommand{\figOneQTwoFreeFrequency}{0.2120}
\newcommand{\figOneQZeroFModeFrequency}{0.2215}
\newcommand{\figOneDifferenceFrequency}{0.1007}
\newcommand{\figOneRadialFrequency}{0.3173}

\hypersetup{
	pdftitle={Nonlinear Tides of Binary Neutron Stars},
	pdfauthor={Guangzhou Guo, Zhen Pan}
}

\begin{document}
	
	\title{Nonlinear Tides of Binary Neutron Stars}
	
	\author{Guangzhou Guo}
	\thanks{{\href{mailto:guangzhou@sjtu.edu.cn}{guangzhou@sjtu.edu.cn}}}
	\affiliation{Tsung-Dao Lee Institute, Shanghai Jiao-Tong University, Shanghai, 520 Shengrong Road, 201210, People’s Republic of China}
	\author{Zhen Pan}
	\thanks{{\href{mailto:zhpan@sjtu.edu.cn}{zhpan@sjtu.edu.cn}}}
	\affiliation{Tsung-Dao Lee Institute, Shanghai Jiao-Tong University, Shanghai, 520 Shengrong Road, 201210, People’s Republic of China}
	\affiliation{School of Physics \& Astronomy, Shanghai Jiao-Tong University, Shanghai, 800 Dongchuan Road, 200240, People’s Republic of China}

	\begin{abstract}
		Nonlinear tides couple stellar oscillations and reshape the
		gravitational wave (GW) signal from inspiralling binaries.
		We present hydrodynamic simulations of nonlinear tides in inspiralling neutron-star binaries and, 
		within a mode-coupling model, derive a semi-analytic quadrupolar fundamental mode-frequency shift through second order in the tidal strength.
		The two approaches agree in the perturbative regime,
		but diverge at stronger tidal fields, where perturbation theory breaks down.
		In the late inspiral, nonlinear tides reduce the fundamental-mode
		frequency by ${\sim}10\%$ for equal-mass binaries and add a GW phase shift of $\mathcal{O}(1)$ radian beyond the linear tidal response.
		These effects make nonlinear tides a promising target for
		next-generation GW detectors.
	\end{abstract}
	
	\maketitle

	\emph{Introduction.}---The first direct detection of gravitational waves (GWs)
	opened a new observational window on compact objects and strong-field gravity
	\cite{Abbott2016,Abbott2016Astrophysics,Abbott2016GR,Abbott2016Properties}.  The binary neutron star (BNS) merger event GW170817 subsequently showed
	that inspiral waveforms can constrain the properties of cold matter at
	densities exceeding those of atomic nuclei, a regime that cannot be reproduced
	in terrestrial laboratories \cite{Abbott2017,Raithel2018,Annala2018,De2018,Abbott2018}.  Future observatories
	will measure BNS signals with substantially greater precision, and such observations are projected to  tighten constraints on stellar radii and the dense-matter Equation of State (EoS) well beyond current limits \cite{Punturo2010,Reitze2019,Maggiore2020,Chatziioannou2020,Pacilio2022,Finstad2023,Huxford2024,Walker2024}.  
	Extracting this information requires waveform models that
	resolve the finite-size dynamics of the stars throughout the late inspiral \cite{DamourNagar2010,Vines2011,Bernuzzi2015,Dietrich2017,Dietrich2019,Gamba2021,PitrePoisson2025,BretzYu2026}.
	
	At low orbital frequencies, tides imprint stellar structure on the GW
	signal through an approximately adiabatic response characterized by the static
	tidal deformability 
	\cite{Lai1994,FlanaganHinderer2008,Hinderer2008,BinningtonPoisson2009,DamourNagar2009,Postnikov2010,DamourNagarVillain2012}.
	At higher orbital frequencies, dynamical tides introduce a frequency-dependent
	response dominated by the quadrupolar fundamental (f-) mode
	\cite{Hinderer2016,SchmidtHinderer2019,AnderssonPnigouras2020,Pratten2020,AnderssonPnigouras2021,Passamonti2021,Williams2022,Pratten2022,GambaBernuzzi2023,PitrePoisson2024,Hegade2024}.
	When the tidal driving frequency sweeps through a stellar
	eigenfrequency, resonant excitation transfers energy from the orbit into the
	mode, modifying the accumulated GW phase \cite{ReiseneggerGoldreich1994,Lai1994,HoLai1999,YuWeinberg2017,XuLai2017,Pan2020,Poisson:2020eki,
		MaYuChen2020,Steinhoff2021,Miao:2023jqe,YuSpin2024,Miao:2025utd,Pnigouras2025,Gao:2025aqo}.
	

Most inspiral models treat this response as  linear.  Near merger, however,
finite tidal deformation changes the restoring forces and couples the f-mode
to other stellar modes \cite{Schenk2002,Weinberg2012,Yu2023,Yu2026}.  Mode coupling analyses
predict a nonlinear shift in the f-mode frequency and an associated orbital phase correction
\cite{Yu2023,Yu2026}. Numerical relativity has recently followed an f-mode resonance  into the nonlinear regime in binaries with highly retrograde spins \cite{Kuan2025}.  In those systems, rotation is what tunes the f-mode frequency into the binary inspiral phase. The off-resonant frequency shift of normally low-spin stars in the inspiral phase, by contrast, requires no such tuning and has not to our knowledge been tested directly against full  hydrodynamic  simulations.


In this work, we study fully nonlinear tides of BNSs
using Newtonian hydrodynamic  simulations and extract mode frequencies by applying Fourier transform to the mass quadrupole moments.
Working within a mode coupling model, we formulate the mode frequency shift in nonlinear regime as an eigenvalue problem,
and derive a semi-analytic expression for the
f-mode frequency shift using perturbation theory through second order
in tidal strength.
We show that the hydrodynamic and mode-coupling frequency shifts agree in the perturbative
regime. 
In the late inspiral, nonlinear tides reduce the f-mode
frequency by ${\sim}10\%$ for equal-mass binaries, and coupling the fluid quadrupole to the orbital motion adds a GW phase shift of 
$\mathcal{O}(1)$ radian beyond the linear tidal response.

\emph{From linear tides to mode coupling.}---For a nonrotating Newtonian star, we expand the displacement
as ${\bm\xi}=\sum_a c_a(t){\bm\xi}_a$, where $c_a$ is a canonical mode
amplitude and ${\bm\xi}_a$ and $\omega_a$ are the eigenfunction
and frequency of the isolated star, with angular indices ${(l,m)}$. In this study, we use the normalization
$2\omega_a^2\int\rho_0|{\bm\xi}_a|^2\,d^3x=E_*=GM_*^2/R_*$,
where $\rho_0$, $M_*$, and $R_*$ are the unperturbed stellar density, mass,
and radius.

The linear tidal response is governed by
\begin{equation}
	\dot c_a+\ii\omega_ac_a=\ii\omega_aU_a.
	\label{eq:linear-response}
\end{equation}
For the quadrupolar tide, only $l=2$ modes are driven, with
$U_a=\epsilon W_{2m}I_ae^{-\ii m\phi}$ and tidal strength
$\epsilon=q(R_*/D)^3$~\cite{Lai1994,ReiseneggerGoldreich1994}.
Here $q=M_c/M_*$ is the ratio of the companion mass to the stellar mass, $D$ and $\phi$ are
the orbital separation and phase, and
$W_{2m}$ are geometric factors, with $W_{20}=-\sqrt{\pi/5}$ and
$W_{2,\pm2}=\sqrt{3\pi/10}$ for an equatorial companion.  The linear tidal overlap is
$I_a=(M_*R_*^2)^{-1}\int\rho_0\,{\bm\xi}_a^*\!\cdot
\bm\nabla(r^2Y_{2m})\,d^3x$, denoted $I_{\rm f}$ for the quadrupolar f-mode.

\begin{figure*}[!t]
	\centering
	\includegraphics[width=0.92\textwidth]{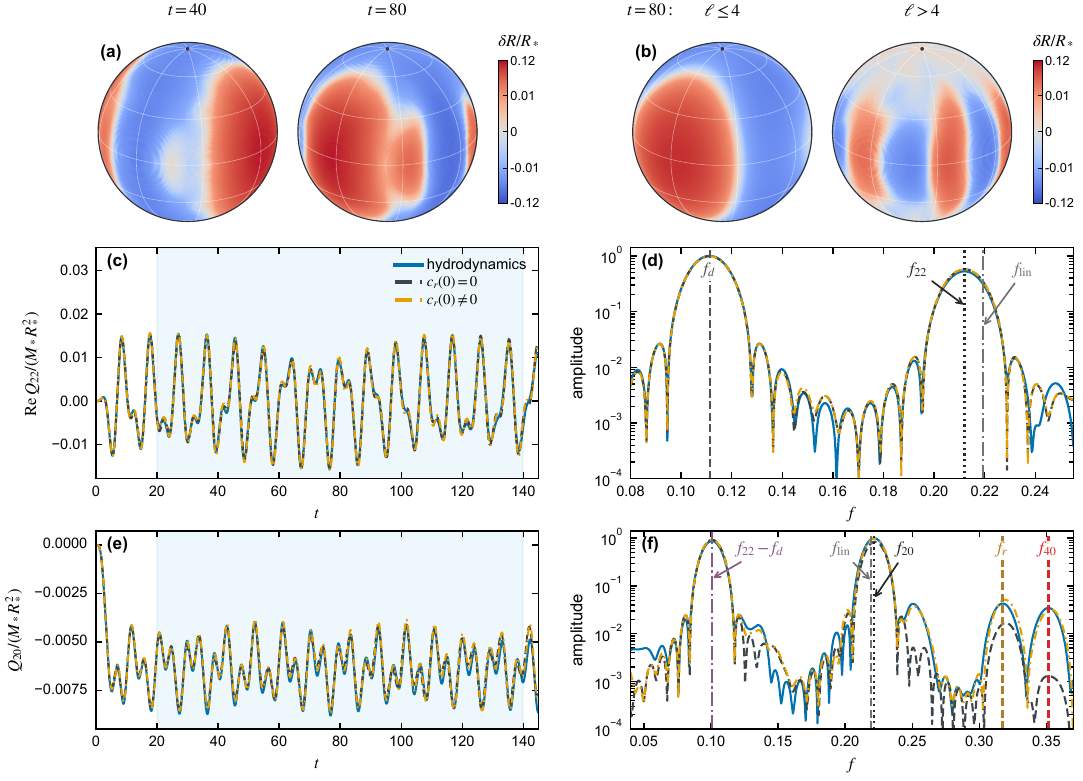}
	\caption{Tidal response for
		$D=\figOneDistance R_*$, $q=0.5$, and $T_{\rm ramp}=5$.
		(a) Tidal deformation at $t=40$ and $80$.
		Colors show the fractional tidal change in surface radius $\delta R/R_*$. 
		(b) The $l\leq4$ and $l>4$ contributions to the tidal deformation at $t=80$.
		For the quadrupoles $Q_{22}$ (middle panels) and $Q_{20}$ (lower panels), hydrodynamic results
		are shown in blue; solutions of the mode coupling equation~\eqref{eq:mode}
		for the $l=0,2,4$ mode set are mapped to $Q_{2m}$
		for $c_r(0)=0$ (dark grey) and
		$c_r(0)\ne0$ (orange).  The shaded interval
		$\figOneWindowStart\leq t\leq\figOneWindowEnd$ is used for  the spectrum analysis.
		The peaks at $f_d$ and $f_{22}$ in $Q_{22}$ correspond to the response driven by the companion and the f-mode of the tidally deformed star, respectively.
		Nonlinear mixing of these two oscillations gives a peak at the difference
		frequency $f_{22}-f_d$ in $Q_{20}$.  The peaks at $f_{20}$, $f_r$, and $f_{40}$ correspond
		to the $(2,0)$ f-mode, the $(0,0)$ mode dominated by radial motion, and the $(4,0)$ f-mode, respectively.}
	\label{fig:modes}
\end{figure*}

In the adiabatic limit, where the binary inspiral timescale is much
longer than the orbital timescale, the off-resonant linear solution is simply
\begin{equation}
	c_a(t)\mathrel{\simeq}\frac{\omega_a U_a(t)}{\omega_a-m\Omega}
	+A_a e^{-\ii\omega_at}.
	\label{eq:linear-tide}
\end{equation}
Here $\Omega=\dot\phi$ is the orbital angular frequency.
For the quadrupolar $m=2$ f-mode, the two terms describe the driven response
at $2\Omega$ and free oscillation at $\omega_f$; $A_a$ is fixed by the
initial condition.  For $|{m}\Omega|\ll|\omega_a|$, the forced component
approaches $U_a$, giving an adiabatic quadrupole response set by
the static tidal deformability~\cite{FlanaganHinderer2008,Hinderer2008}.
The induced quadrupole changes the orbital evolution and accumulated GW
phase~\cite{Lai1994,Hinderer2016}.  In this linear approximation, the tide produces a
forced response without shifting the mode frequencies of the star.

Beyond linear response, the coupling between different modes follows a perturbative expansion of the stellar dynamics~\cite{Schenk2002,Weinberg2012,Yu2023},
\begin{multline}
	\dot c_a+\ii\omega_ac_a=\ii\omega_a\Bigl(
	U_a+\sum_bU_{ab}^*c_b^*+\sum_{bc}\kappa_{abc}c_b^*c_c^* \\
	{}+\sum_{bcd}\zeta_{abcd}c_b^*c_c^*c_d^*\Bigr),
	\label{eq:mode}
\end{multline}
where $U_{ab}= \epsilon\sum_{m=0,\pm2}W_{2m}J_{ab;2m}e^{-\ii m\phi}$
describes nonlinear tidal driving.  The quadratic overlap integral
$J_{ab;2m}=(M_*R_*^2)^{-1}\int\rho_0\,{\bm\xi}_a\!\cdot
({\bm\xi}_b\!\cdot\bm\nabla)\bm\nabla(r^2Y_{2m})\,d^3x$
couples modes $a$ and $b$ with the tide~\cite{Yu2023}.  The coefficients $\kappa_{abc}$ and
$\zeta_{abcd}$ describe three- and four-mode interactions within the star (see Supplemental Material for definitions).

Mode couplings excite new modes that the tide does not drive directly, e.g., 
quadratic coupling between $c_{22}$ and
$c_{2,-2}$ excites modes with $m=0$~\cite{Schenk2002,Weinberg2012} (see Fig.~\ref{fig:modes}), 
\begin{equation}
	c_{22}\otimes c_{2,-2}\ \longrightarrow\
	c_r\oplus c_{20}\oplus c_{40}\ ,
	\label{eq:mzero-channels}
\end{equation}
where $c_r$ is the $(0,0)$ radial mode amplitude.
Such couplings shift the mode frequencies and thereby modify the accumulated GW
phase through changes in the tidal response~\cite{Yu2023}.

\emph{Fully nonlinear tides.}---We evolve a nonrotating,
self-gravitating $\Gamma=2$ polytrope in Newtonian gravity with
Athena++~\cite{Stone2020}.  The unperturbed star has $M_*=2.5066$,
$R_*=1.2534$, and the quadrupolar f-mode frequency $f_{\rm lin}=\omega_f/(2\pi)=\fAthena$ in units $G=K=\rho_c=1$.  
We solve the nonlinear Euler--Poisson equations
directly, without  truncation in mode amplitudes,
\begin{subequations}\label{eq:hydrodynamics}
	\begin{align}
		\partial_t\rho+\bm\nabla\!\cdot(\rho\bm v)&=0,\\
		\partial_t(\rho\bm v)+\bm\nabla\!\cdot
		(\rho\bm v\bm v+P\bm I)&=-\rho\bm\nabla(\Phi+U_{\rm tide}),\\
		\partial_tE+\bm\nabla\!\cdot[(E+P)\bm v]
		&=-\rho\bm v\!\cdot\bm\nabla(\Phi+U_{\rm tide}),\\
		\nabla^2\Phi&=4\pi G\rho,
	\end{align}
\end{subequations}
Here $\rho$, $\bm v$, $P$, $E$, and $\Phi$ are the density, velocity,
pressure, total fluid energy density, and self-gravitational potential.  The
EoS is $P=(\Gamma-1)(E-\rho v^2/2)$.  The companion is represented by the prescribed
quadrupolar tidal potential
\begin{equation}
	U_{\rm tide}(\bm x,t)=-\frac{GM_*}{R_*^3}\,\epsilon(t)\,r^2
	\sum_{m=0,\pm2}W_{2m}Y_{2m}(\hat{\bm x})e^{-\ii m\phi},
	\label{eq:tide}
\end{equation}
with $\hat{\bm x}=\bm x/r$.  The time-dependent tidal strength is
$\epsilon(t)=g(t)q[R_*/D(t)]^3$, where the envelope $g(t)=\sin^2[\pi t/(2T_{\rm ramp})]$ for
$0\leq t<T_{\rm ramp}$ smoothly raises the tidal field from zero and $g(t)=1$ for $t\geq T_{\rm ramp}$.

In Fig.~\ref{fig:modes}, we show the tidal response of a neutron star for
$D=\figOneDistance R_*$ and $q=0.5$.
Panel (a) exhibits the surface deformation
$\delta R/R_*=(R_{\rm s}-R_*)/R_*$, where $R_{\rm s}(\theta, \phi)$ is the deformed surface radius bounding the region with $\rho\ge 10^{-3}\rho_c$. Decomposing $\delta R/R_*$ at $t=80$ into spherical harmonics separates the $l\leq4$ and $l>4$
contributions shown in Fig.~\ref{fig:modes}(b).
The hydrodynamic evolution also captures weaker $l>4$ contributions,
implying higher order interactions between tidal modes.

To extract tidal mode information, we compute the mass quadrupole moments
$Q_{2m}(t)=\int_{\mathcal V_*}\rho r^2Y_{2m}^*d^3x$ on circular orbits at fixed
$D$ and $\Omega$ after the ramp, and obtain their spectra by fast Fourier
transform.  Fig.~\ref{fig:modes}(c)--(f) show the quadrupole moments and their spectra.
The orbital angular frequency follows
the Kepler's law,
$\Omega=[GM_*(1+q)/D^3]^{1/2}=\figOneOrbitalAngularFrequency$, giving $m=2$
forcing at $f_d=2\Omega/(2\pi)=\figOneDriveFrequency$.  The $Q_{22}$ peak driven by the companion occurs exactly at $f_d$, while the tidal deformation
shifts the f-mode from $f_{\rm lin}=0.2194$ to
$f_{22}=\figOneQTwoFreeFrequency$.
The $Q_{20}$ spectrum contains several
additional peaks, which we interpret below.

\emph{Mode coupling interpretation.}---To interpret the hydrodynamic spectra, we solve
Eq.~\eqref{eq:mode} for the $l=0,2,4$ mode set ($X_{lm}\equiv c_{lm}+c_{l,-m}^{*}$,
$X_r\equiv c_r+c_r^{*}$) and map the amplitudes to
mass quadrupole moments at quadratic order~\cite{Yu2026}, 
\begin{align}
	\frac{Q_{22}}{M_*R_*^2}
	&=(I_f+J_2X_{20}+J_{2r}X_r)X_{22}+\Delta_{22}^{(4)},
	\label{eq:quadrupole-modes}\\
	\frac{Q_{20}}{M_*R_*^2}
	&=\left(I_f-\tfrac12J_2X_{20}+J_{2r}X_r\right)X_{20}
	+J_2|X_{22}|^2+\Delta_{20}^{(4)}.\nonumber
\end{align}
where $J_2$ and $J_{2r}$ are the f--f and f--radial quadratic overlaps,
respectively. Here
$\Delta_{2m}^{(4)}=\sum_{m_1+m_2=m}J_{(2,m_1)(4,m_2);2,-m}X_{2m_1}X_{4m_2}
+\tfrac12\sum_{m_1+m_2=m}J_{(4,m_1)(4,m_2);2,-m}X_{4m_1}X_{4m_2}$ marks the contribution of $(4,m)$ modes via nonlinear coupling.   


Panels (c) and (e) of Fig.~\ref{fig:modes} show the waveforms
reconstructed from the mode coupling solution, along with the
hydrodynamic simulation results \footnote{When
	solving Eq.~\eqref{eq:mode}, $c_r(0)$ sets the initial
	amplitude of the $l=0$ radial mode.  Fig.~\ref{fig:modes} uses the
	$T_{\rm ramp}=5$ hydrodynamic simulation.  For the $c_r(0)\ne0$ mode coupling
	solution, we determine the complex initial radial mode amplitude by fitting
	the amplitude and phase of the radial peak in an independent
	$T_{\rm ramp}=20$ hydrodynamic simulation.  This added homogeneous component
	mimics the enhanced radial mode excitation caused by numerical perturbations
	when the spherical equilibrium is mapped onto the numerical mesh.}.
As shown in panels (d) and (f) of Fig.~\ref{fig:modes}, the mode coupling model captures the main spectral features.  In the quadratic term of Eq.~\eqref{eq:quadrupole-modes}, the
component of $Q_{22}$ driven by the companion at
$f_d=\figOneDriveFrequency$ and the tidally shifted f-mode component at $f_{22}=\figOneQTwoFreeFrequency$ interact to
give rise to a peak at the difference frequency $f_{22}-f_d=\figOneDifferenceFrequency$ in the
$Q_{20}$ spectrum.

The $(0,0)$ radial mode can be excited by mode coupling and initial
perturbations.  In a spherical star, radial oscillations do not contribute to the
quadrupole moment at linear order.  Tidal deformation breaks this symmetry,
allowing radial oscillations to change $Q_{20}$ through the nonlinear
$c_{20}$--$c_r$ coupling in Eq.~\eqref{eq:quadrupole-modes}.  The resulting peak in
the $Q_{20}$ spectrum occurs at $f_r=\figOneRadialFrequency$, 8.8\% below the isolated star frequency $f_{r,{\rm iso}}=0.3480$ \footnote{The value of $f_{r,{\rm iso}}$ is also
	checked by a hydrodynamic simulation without tides, which gives a radial mode peak at
	$0.3457$ (within 0.7\% of GYRE~\cite{Townsend2013}).}.
In addition,  the $(4,0)$ f-mode is also excited at
$f_{40}=0.3515$, which is $3.9\%$ below the isolated-star value
$f_{40,{\rm iso}}=0.3657$~\cite{Townsend2013}.
The remaining peak at
$f_{20}=\figOneQZeroFModeFrequency$ is the
nonlinearly excited $(2,0)$ f-mode, whose tidal frequency shift is less
pronounced than that of the $(2,2)$ f-mode.

\begin{figure}[!t]
	\centering
	\includegraphics[width=\columnwidth]{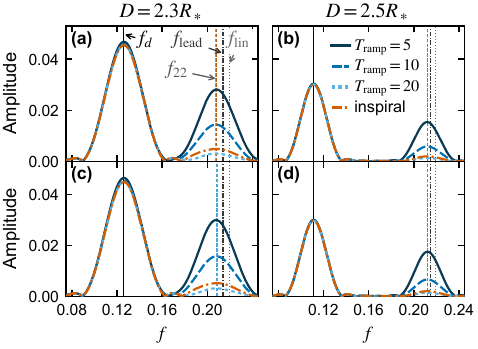}
	\caption{$Q_{22}$ amplitude spectra for $q=0.5$ at $D/R_*=2.3$ and $2.5$.
		The upper row shows hydrodynamic simulations; the lower row shows solutions of
		Eq.~\eqref{eq:mode} for the $l=0,2,4$ mode coupling model.  Evolutions
		at fixed separation use $T_{\rm ramp}=5,10,20$ (dark to light blue).  For the
		data prepared through inspiral, $D$ decreases from $10R_*$ to the target separation by
		$t=50$.  Vertical lines mark $f_d$, $f_{\rm lin}$,
		$f_{\rm lead}=f_{\rm lin}+\Delta f^{(1)}$, and $f_{22}$.  Varying the f-mode
		excitation changes the peak amplitude but not its nonlinear frequency $f_{22}$.}
	\label{fig:ramp}
\end{figure}

\emph{Semi-analytic formula for mode frequency shifts.}---To understand how the
quadrupolar $m=2$ f-mode shifts from the frequency $f_{\rm lin}$ of the isolated star
to the frequency $f_{22}$ of the tidally deformed star, we compute the first- and
second-order frequency shifts perturbatively.  We first construct the stationary tidal
solution of Eq.~\eqref{eq:mode} and then linearize small $m=2$ oscillations
on that background.

At fixed $D$ and $\Omega$, we remove the orbital phase by defining
$b_{lm}=e^{\ii m\phi}c_{lm}$.  Collecting the transformed amplitudes and their
conjugates in $\bm C$ gives the time-independent system
\begin{equation}
	\dot{\bm C}=\bm F(\bm C;\epsilon).
	\label{eq:autonomous-system}
\end{equation}
The stationary tide satisfies $\bm F(\overline{\bm C};\epsilon)=0$.
Linearizing Eq.~\eqref{eq:autonomous-system} about this equilibrium defines
$\mathsf J(\epsilon)\equiv
\left.\partial\bm F/\partial\bm C\right|_{\overline{\bm C}}$.  A perturbation
$\delta\bm C=\bm v e^{-\ii\sigma t}$ then obeys
\begin{equation}
	\mathsf J(\epsilon)\bm v=-\ii\sigma\bm v.
	\label{eq:linearized-mode}
\end{equation}
Eq.~\eqref{eq:linearized-mode} is the eigenvalue problem for small
oscillations about the tidal equilibrium.  For the eigenvector $\bm v_0$
associated with the $m=2$ f-mode, the frequency in the inertial frame satisfies
$2\pi f_{22}=\sigma+2\Omega$, where the $2\Omega$ term restores the orbital
phase removed in Eq.~\eqref{eq:autonomous-system}.  At $\epsilon=0$, denote
the right eigenvectors by $\bm v_\nu^{(0)}$ and choose the left eigenvectors
$\bm l_\nu^{(0)}$ so that
$\bm l_\nu^{(0)\dagger}\mathsf J^{(0)}
=-\ii\sigma_\nu^{(0)}\bm l_\nu^{(0)\dagger}$ and
$\bm l_\mu^{(0)\dagger}\bm v_\nu^{(0)}=\delta_{\mu\nu}$; $\nu=0$ denotes
the $m=2$ f-mode of the isolated star.  A series expansion of
Eq.~\eqref{eq:linearized-mode},
$\mathsf J=\mathsf J^{(0)}+\epsilon\mathsf J^{(1)}
+\epsilon^2\mathsf J^{(2)}+O(\epsilon^3)$ and
$\bm v_0=\bm v_0^{(0)}+\epsilon\bm v_0^{(1)}+O(\epsilon^2)$, gives
$f_{22}=f_{\rm lin}+\Delta f^{(1)}+\Delta f^{(2)}+O(\epsilon^3)$, where
\begin{equation}
	\begin{aligned}
		\frac{2\pi\Delta f^{(1)}}{\epsilon}
		&=\ii\bm l_0^{(0)\dagger}\mathsf J^{(1)}\bm v_0^{(0)},\\
		\frac{2\pi\Delta f^{(2)}}{\epsilon^2}
		&=\ii\bm l_0^{(0)\dagger}\mathsf J^{(2)}\bm v_0^{(0)}
		+\ii\bm l_0^{(0)\dagger}\mathsf J^{(1)}\bm v_0^{(1)}.
	\end{aligned}
	\label{eq:eigenvector-frequency-shifts}
\end{equation}
Here the leading-order correction to the $m=2$ f-mode eigenvector is
$\bm v_0^{(1)}=\ii\sum_{\nu\ne0}
(\bm l_\nu^{(0)\dagger}\mathsf J^{(1)}\bm v_0^{(0)})\allowbreak
\bm v_\nu^{(0)}/(\sigma^{(0)}-\sigma_\nu^{(0)})$, with
$\bm l_0^{(0)\dagger}\bm v_0^{(1)}=0$.
Evaluating Eq.~\eqref{eq:eigenvector-frequency-shifts} explicitly, as done in
the Supplemental Material, gives the semi-analytic perturbative shift of $f_{22}$.  Notably,
the shift depends only on the tidally deformed equilibrium $\overline{\bm C}$ and not on the
amplitude of the excited eigenstate $\bm v_0$.  Thus, within this
small-oscillation approximation, changing the excitation amplitude does not
shift $f_{22}$.

This amplitude independence is directly testable in the hydrodynamic simulations.
We displace the star from its tidal equilibrium to excite the f-mode, using several preparations of the same target orbit $D=2.3 R_*/2.5 R_*$: at fixed separation we take
$T_{\rm ramp}=5$, $10$, and $20$, and we also prepare the star through an
inspiral in which $D$ decreases from $10R_*$ to the final separation by
$t=50$.
As shown in the upper row of Fig.~\ref{fig:ramp}, a shorter
$T_{\rm ramp}$ excites a larger f-mode amplitude, while all preparations yield
the same $f_{22}$, marked by the common vertical line.  The lower row shows
the corresponding $l=0,2,4$ mode-coupling solutions, which reproduce both the
variation in peak amplitude and the invariance of $f_{22}$.

\begin{figure}[!t]
	\centering
	\includegraphics[width=\columnwidth]{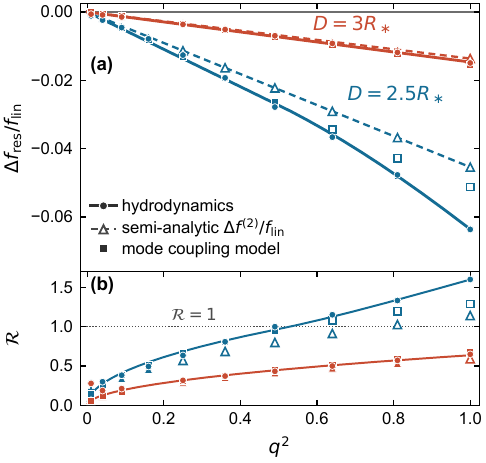}
	\caption{(a) Residual f-mode shift $\Delta f_{\rm res}/f_{\rm lin}$, with $\Delta f_{\rm res}=f_{22}-(f_{\rm lin}+\Delta f^{(1)})$, against $q^2$ for two orbital scenarios at
		$(D/R_*,\Omega)=(2.5,\omegaTwoFive)$ and $(3,\omegaThree)$; 
		blue and red denote the two scenarios.  Circles and squares show hydrodynamic
		results and the time-dependent $l=0,2,4$ mode coupling model
		in Eq.~\eqref{eq:mode}.  Triangles show the semi-analytic
		$\Delta f^{(2)}/f_{\rm lin}$ from
		Eq.~\eqref{eq:eigenvector-frequency-shifts}.
		Solid curves are hydrodynamic fits, while dashed lines show the
		semi-analytic $q^2$ scaling.
		(b) $\mathcal R=|\Delta f_{\rm res}/\Delta f^{(1)}|$ for all three
		approaches, using $\Delta f^{(2)}$ for the semi-analytic residual.
		The dotted line marks $\mathcal R=1$.}
	\label{fig:frequency}
\end{figure}

To determine how the nonlinear frequency correction scales with tidal
strength $\epsilon$, we consider two orbital scenarios and vary $q$ while holding $D$
and $\Omega$ fixed, with $0\le q\le1$.
At fixed $D$, $\epsilon\propto q$, so the second-order tidal correction
scales as $q^2$.  We remove the leading contribution
$\Delta f^{(1)}$ by defining
$f_{\rm lead} := f_{\rm lin}+\Delta f^{(1)}$.  The remaining frequency
shift,
$\Delta f_{\rm res}:= f_{22}-f_{\rm lead}$, then equals $\Delta f^{(2)}$ up
to $\mathcal{O}(\epsilon^3)$.  We quantify the perturbative
ordering by
$\mathcal R :=|\Delta f_{\rm res}/\Delta f^{(1)}|$.  For the hydrodynamic and time-dependent mode coupling results,
Fig.~\ref{fig:frequency}(b) shows that $\mathcal R<1$ up to
$q^2\simeq0.49$ at $D=2.5R_*$ and throughout the sampled
range at $D=3R_*$.  In this regime, the filled circles and
squares in Fig.~\ref{fig:frequency}(a) show consistency between
hydrodynamics and the mode coupling model.  Once
$\mathcal R\geq1$, the quadratic correction is no longer subleading to
$\Delta f^{(1)}$, and a second-order truncation is not perturbatively
controlled.  The open squares show deviations from
hydrodynamics at stronger tides.

In Fig.~\ref{fig:frequency}(a), the dashed lines denoting the
semi-analytic $\Delta f^{(2)}$ follow the $q^2$ scaling by construction.
Their agreement with the hydrodynamic results at weaker tides tests the expansion of
Eq.~\eqref{eq:linearized-mode} through $\mathcal{O}(\epsilon^2)$,
while the stronger-tide results depart from this quadratic prediction.
In the late inspiral $D=2.5 R_*$, where the second-order truncation is no longer controlled, nonlinear tides reduce the f-mode frequency by $\sim 10\%$ for equal-mass binaries.


\emph{GW phase shift from nonlinear tides.}---
The tidally induced stellar quadrupole adds a contribution to the
binary interaction, thereby altering the orbital frequency evolution and the
accumulated GW phase.  In the adiabatic
limit, the quadrupole follows the tidal field and the phase is governed by the
tidal deformability \cite{FlanaganHinderer2008,Hinderer2010}.  As the tidal
driving frequency $2\Omega$ approaches the f-mode frequency $\omega_f$, the finite response time of the
star generates the dynamical tide correction
\cite{Hinderer2016,Steinhoff2016,SchmidtHinderer2019}.
Low-order stellar modes also couple to the quadrupolar response
\cite{Schenk2002,Weinberg2012};  such couplings reduce the f-mode frequency,
strengthen its finite-frequency response, and generate an additional GW 
phase beyond linear tidal response \cite{Yu2023}. 



To quantify the GW phase shift from fully nonlinear tides, we evolve an equal-mass $1.4 M_\odot+1.4M_\odot$ inspiral.
The stellar parameters are $GM_*/(R_*c^2)=0.17$, $R_*=12.1\,\mathrm{km}$, and
$f_{\rm lin}=1976\,\mathrm{Hz}$.  The tidal potential of
Eq.~\eqref{eq:tide} is turned on slowly with $T_{\rm ramp}=20$ at $D=5R_*$,
at which point $f_{\rm GW}\simeq400$ Hz; we then evolve the coupled
quadrupole--orbit system up to 1100 Hz, where quadrupole moments $Q_{20}(t)$ and $Q_{22}(t)$ are obtained by evolving hydrodynamic Eqs.~(\ref{eq:hydrodynamics}). The quadrupole couples to the orbit through the interaction energy
\begin{equation}
	H_{\rm tide}=-\frac{GN_*M_c}{D^3}
	\left[W_{20}Q_{20}+2W_{22}\operatorname{Re}
	\left(e^{2\ii\phi}Q_{22}\right)\right],
	\label{eq:tidal-hamiltonian}
\end{equation}
where $N_*=2$ counts the two identically deformed stars, $M_c$ is the companion mass and $Q_{20}$ and $Q_{22}$ are the mass
quadrupole moments defined above.  With reduced mass
$\mu=M_*M_c/(M_*+M_c)$, this energy perturbs the relative orbit, giving the leading-order
radial and tangential tidal accelerations
$a_D^{\rm tide}=-(1/\mu)\partial_DH_{\rm tide}$ and
$a_\phi^{\rm tide}=-(1/\mu D)\partial_\phi H_{\rm tide}$ at fixed stellar
quadrupole:
\begin{align}
	a_D^{\rm tide}&=-\frac{6GN_*(1+q)}{D^4}
	\left[W_{22}\operatorname{Re}\left(e^{2\ii\phi}Q_{22}\right)
	+\frac{W_{20}}{2}Q_{20}\right],\nonumber\\
	a_\phi^{\rm tide}&=-\frac{4GN_*(1+q)W_{22}}{D^4}
	\operatorname{Im}\left(e^{2\ii\phi}Q_{22}\right).
	\label{eq:tidal-accelerations}
\end{align}
The real and imaginary parts of $e^{2\ii\phi}Q_{22}$ describe the deformation
aligned with the companion and the component responsible for the tidal torque,
respectively~\cite{Lai1994,Yu2023}.  We evolve the orbital elements with these accelerations and the leading-order radiation reaction (see Supplemental Material for more details).


\begin{figure}[!t]
	\centering
	\includegraphics[width=\columnwidth]{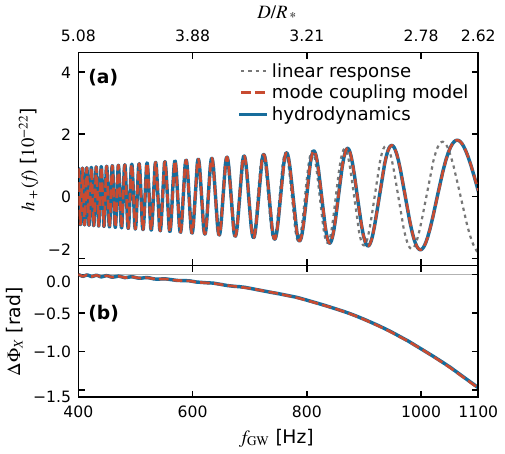}
	\caption{GW signal from an equal-mass $1.4 M_\odot+1.4M_\odot$ neutron star binary.
		(a) GW strain $h_+$ for a face-on observer at 100 Mpc,
		computed from the orbital and stellar mass quadrupoles.
		Grey dotted, red dashed, and blue solid curves denote the linear dynamical
		tidal response, the $l=0,2,4$ mode coupling model, and hydrodynamics.
		(b) Phase shifts $\Delta\Phi_X$ relative to the linear dynamical tidal
		response for hydrodynamics and the mode coupling model.}
	\label{fig:phase}
\end{figure}

For these evolutions, we define the
dominant $m=2$ GW phase shift relative to the linear dynamical tidal response as
\begin{equation}
	\Delta\Phi_X(f)=2\bigl(\phi_X(f)-\phi_{\rm lin}(f)\bigr),
	\label{eq:phase-shift}
\end{equation}
where $\phi_X(f)$ is the orbital phase at GW frequency
$f=\Omega/\pi$ for the mode coupling or hydrodynamic
evolution, and $\phi_{\rm lin}(f)$ is that of the linear
dynamical tidal response.
A negative $\Delta\Phi_X$ indicates that the binary reaches a given GW
frequency after accumulating less phase than in the linear response.
As the inspiral proceeds, the hydrodynamic and mode coupling
waveforms in Fig.~\ref{fig:phase}(a) dephase from the linear response.
The corresponding phase shifts $\Delta\Phi_X$ are shown in
Fig.~\ref{fig:phase}(b).
By 1100 Hz, both calculations differ from the linear response by
$\mathcal{O}(1)$ radian, consistent with previous estimates of nonlinear tides \cite{Yu2023}.

\emph{Discussion.}--- We presented hydrodynamic simulations of
nonlinear tides in inspiralling BNSs, extracting the nonlinear-tide-induced
f-mode frequency shift and GW phase shift.  To understand the frequency shift
in the nonlinear regime, we formulated it as an eigenvalue problem within the
mode-coupling model and derived a semi-analytic expression through second order
in the tidal strength.  The consistency of the two approaches in the
perturbative regime supports the mode-coupling model as a quantitative
description of the nonlinear f-mode frequency shift, while at stronger tides
the hydrodynamic results probe the limits of the perturbative description.  In
the late inspiral, nonlinear tides reduce the f-mode frequency by ${\sim}10\%$
for equal-mass binaries.  By 1100 Hz, the quadrupole--orbit coupling adds a GW
phase shift of $\mathcal{O}(1)$ radian beyond the linear tidal response.
These effects make nonlinear tides a promising target for next-generation GW
detectors~\cite{Williams2022}.

The linear tidal response and its application to probing the neutron-star
EoS have been studied extensively; nonlinear tides may play a
complementary role.  In future work, we plan to apply our combined hydrodynamic
and binary-evolution framework to other nonlinear tidal mechanisms, such as
coupling between pressure and gravity modes
\cite{Weinberg2013,Venumadhav2014,Weinberg2016}, and to explore the role of
nonlinear tides in constraining the neutron-star EoS.  
A relativistic extension could test mode coupling predictions against numerical relativity and guide nonlinear tidal prescriptions in waveform models~\cite{Kuan2025}.

\emph{Acknowledgments.}---We are grateful to Peng Wang  for helpful
discussions and valuable
comments.  Guangzhou Guo acknowledges support from the China Postdoctoral
Science Foundation (Grant No.~2025M783414) and the National Natural Science
Foundation of China (Grant No.~12605116).

\clearpage
\renewcommand{\bibfont}{\footnotesize}
\bibliographystyle{apsrev4-2}
\bibliography{nonlinear_tidal_modes}

@article{Abbott2016Astrophysics,
	author = {{LIGO Scientific Collaboration} and {Virgo Collaboration}},
	title = {Astrophysical Implications of the Binary Black Hole Merger {GW150914}},
	journal = {Astrophys. J. Lett.},
	volume = {818},
	pages = {L22},
	year = {2016},
	doi = {10.3847/2041-8205/818/2/L22}
}

@article{Abbott2016GR,
	author = {{LIGO Scientific Collaboration} and {Virgo Collaboration}},
	title = {Tests of General Relativity with {GW150914}},
	journal = {Phys. Rev. Lett.},
	volume = {116},
	pages = {221101},
	year = {2016},
	doi = {10.1103/PhysRevLett.116.221101},
	note = {Erratum: Phys. Rev. Lett. 121, 129902 (2018)}
}

@article{Abbott2016Properties,
	author = {{LIGO Scientific Collaboration} and {Virgo Collaboration}},
	title = {Properties of the Binary Black Hole Merger {GW150914}},
	journal = {Phys. Rev. Lett.},
	volume = {116},
	pages = {241102},
	year = {2016},
	doi = {10.1103/PhysRevLett.116.241102}
}

@article{De2018,
	author = {De, Soumi and Finstad, Daniel and Lattimer, James M. and Brown, Duncan A. and Berger, Edo and Biwer, Christopher M.},
	title = {Tidal Deformabilities and Radii of Neutron Stars from the Observation of {GW170817}},
	journal = {Phys. Rev. Lett.},
	volume = {121},
	pages = {091102},
	year = {2018},
	doi = {10.1103/PhysRevLett.121.091102},
	note = {Erratum: Phys. Rev. Lett. 121, 259902 (2018)}
}

@article{Annala2018,
	author = {Annala, Eemeli and Gorda, Tyler and Kurkela, Aleksi and Vuorinen, Aleksi},
	title = {Gravitational-Wave Constraints on the Neutron-Star-Matter Equation of State},
	journal = {Phys. Rev. Lett.},
	volume = {120},
	pages = {172703},
	year = {2018},
	doi = {10.1103/PhysRevLett.120.172703}
}

@article{Raithel2018,
	author = {Raithel, Carolyn A. and {\"O}zel, Feryal and Psaltis, Dimitrios},
	title = {Tidal Deformability from {GW170817} as a Direct Probe of the Neutron Star Radius},
	journal = {Astrophys. J. Lett.},
	volume = {857},
	pages = {L23},
	year = {2018},
	doi = {10.3847/2041-8213/aabcbf}
}

@article{Pacilio2022,
	author = {Pacilio, Costantino and Maselli, Andrea and Fasano, Margherita and Pani, Paolo},
	title = {Ranking Love Numbers for the Neutron Star Equation of State: The Need for Third-Generation Detectors},
	journal = {Phys. Rev. Lett.},
	volume = {128},
	pages = {101101},
	year = {2022},
	doi = {10.1103/PhysRevLett.128.101101}
}

@article{DamourNagar2010,
	author = {Damour, Thibault and Nagar, Alessandro},
	title = {Effective one body description of tidal effects in inspiralling compact binaries},
	journal = {Phys. Rev. D},
	volume = {81},
	pages = {084016},
	year = {2010},
	doi = {10.1103/PhysRevD.81.084016}
}

@article{Postnikov2010,
	author = {Postnikov, Sergey and Prakash, Madappa and Lattimer, James M.},
	title = {Tidal Love numbers of neutron and self-bound quark stars},
	journal = {Phys. Rev. D},
	volume = {82},
	pages = {024016},
	year = {2010},
	doi = {10.1103/PhysRevD.82.024016}
}

@article{DamourNagarVillain2012,
	author = {Damour, Thibault and Nagar, Alessandro and Villain, Lo{\"i}c},
	title = {Measurability of the tidal polarizability of neutron stars in late-inspiral gravitational-wave signals},
	journal = {Phys. Rev. D},
	volume = {85},
	pages = {123007},
	year = {2012},
	doi = {10.1103/PhysRevD.85.123007}
}

@article{AnderssonPnigouras2020,
	author = {Andersson, Nils and Pnigouras, Pantelis},
	title = {Exploring the effective tidal deformability of neutron stars},
	journal = {Phys. Rev. D},
	volume = {101},
	pages = {083001},
	year = {2020},
	doi = {10.1103/PhysRevD.101.083001}
}

@article{Passamonti2021,
	author = {Passamonti, Andrea and Andersson, Nils and Pnigouras, Pantelis},
	title = {Dynamical tides in neutron stars: the impact of the crust},
	journal = {Mon. Not. R. Astron. Soc.},
	volume = {504},
	number = {1},
	pages = {1273--1293},
	year = {2021},
	doi = {10.1093/mnras/stab870}
}

@article{PitrePoisson2024,
	author = {Pitre, Tristan and Poisson, Eric},
	title = {General relativistic dynamical tides in binary inspirals without modes},
	journal = {Phys. Rev. D},
	volume = {109},
	pages = {064004},
	year = {2024},
	doi = {10.1103/PhysRevD.109.064004}
}

@article{GambaBernuzzi2023,
	author = {Gamba, Rossella and Bernuzzi, Sebastiano},
	title = {Resonant tides in binary neutron star mergers: Analytical-numerical relativity study},
	journal = {Phys. Rev. D},
	volume = {107},
	pages = {044014},
	year = {2023},
	doi = {10.1103/PhysRevD.107.044014}
}

@article{ReiseneggerGoldreich1994,
	author = {Reisenegger, Andreas and Goldreich, Peter},
	title = {Excitation of neutron star normal modes during binary inspiral},
	journal = {Astrophys. J.},
	volume = {426},
	pages = {688--691},
	year = {1994},
	doi = {10.1086/174105}
}

@article{YuWeinberg2017,
	author = {Yu, Hang and Weinberg, Nevin N.},
	title = {Resonant tidal excitation of superfluid neutron stars in coalescing binaries},
	journal = {Mon. Not. R. Astron. Soc.},
	volume = {464},
	number = {3},
	pages = {2622--2637},
	year = {2017},
	doi = {10.1093/mnras/stw2552}
}

@article{Pnigouras2025,
	author = {Pnigouras, P. and Andersson, N. and Gittins, F. and Counsell, A. R.},
	title = {Dynamical neutron star tides: the signature of a mode resonance},
	journal = {Mon. Not. R. Astron. Soc.},
	volume = {542},
	number = {2},
	pages = {1375--1387},
	year = {2025},
	doi = {10.1093/mnras/staf1285}
}

@article{MaYuChen2020,
	author = {Ma, Sizheng and Yu, Hang and Chen, Yanbei},
	title = {Excitation of {f}-modes during mergers of spinning binary neutron star},
	journal = {Phys. Rev. D},
	volume = {101},
	pages = {123020},
	year = {2020},
	doi = {10.1103/PhysRevD.101.123020}
}

@article{XuLai2017,
	author = {Xu, Wenrui and Lai, Dong},
	title = {Resonant tidal excitation of oscillation modes in merging binary neutron stars: Inertial-gravity modes},
	journal = {Phys. Rev. D},
	volume = {96},
	pages = {083005},
	year = {2017},
	doi = {10.1103/PhysRevD.96.083005}
}

@article{Abbott2016,
	author = {{LIGO Scientific Collaboration} and {Virgo Collaboration}},
	title = {Observation of Gravitational Waves from a Binary Black Hole Merger},
	journal = {Phys. Rev. Lett.},
	volume = {116},
	pages = {061102},
	year = {2016},
	doi = {10.1103/PhysRevLett.116.061102}
}

@article{Abbott2017,
	author = {{LIGO Scientific Collaboration} and {Virgo Collaboration}},
	title = {GW170817: Observation of Gravitational Waves from a Binary Neutron Star Inspiral},
	journal = {Phys. Rev. Lett.},
	volume = {119},
	pages = {161101},
	year = {2017},
	doi = {10.1103/PhysRevLett.119.161101}
}

@article{Punturo2010,
	author = {Punturo, M. and others},
	title = {The {Einstein Telescope}: a third-generation gravitational wave observatory},
	journal = {Class. Quantum Grav.},
	volume = {27},
	pages = {194002},
	year = {2010},
	doi = {10.1088/0264-9381/27/19/194002}
}

@article{Miao:2023jqe,
	author = "Miao, Zhiqiang and Zhou, Enping and Li, Ang",
	title = "{Resolving Phase Transition Properties of Dense Matter through Tidal-excited g-mode from Inspiralling Neutron Stars}",
	eprint = "2305.08401",
	archivePrefix = "arXiv",
	primaryClass = "nucl-th",
	doi = "10.3847/1538-4357/ad27cd",
	journal = "Astrophys. J.",
	volume = "964",
	number = "1",
	pages = "31",
	year = "2024"
}

@article{Gao:2025aqo,
	author = "Gao, Yong and Kuan, Hao-Jui and Xia, Cheng-Jun and Silva, Hector O. and Shibata, Masaru",
	title = "{Nonradial oscillations of stratified neutron stars with solid crusts: Mode characterization and tidal resonances in coalescing binaries}",
	eprint = "2509.00257",
	archivePrefix = "arXiv",
	primaryClass = "astro-ph.HE",
	doi = "10.1103/nbk7-8kts",
	journal = "Phys. Rev. D",
	volume = "112",
	number = "12",
	pages = "123006",
	year = "2025"
}

@article{Miao:2025utd,
	author = "Miao, Zhiqiang and Feng, Xuefeng and Pan, Zhen and Yang, Huan",
	title = "{Relativistic excitation of compact stars}",
	eprint = "2506.23176",
	archivePrefix = "arXiv",
	primaryClass = "gr-qc",
	doi = "10.1103/ddjt-ftws",
	journal = "Phys. Rev. D",
	volume = "112",
	number = "10",
	pages = "104054",
	year = "2025"
}

@article{Poisson:2020eki,
	author = "Poisson, Eric",
	title = "{Gravitomagnetic tidal resonance in neutron-star binary inspirals}",
	eprint = "2003.10427",
	archivePrefix = "arXiv",
	primaryClass = "gr-qc",
	doi = "10.1103/PhysRevD.101.104028",
	journal = "Phys. Rev. D",
	volume = "101",
	number = "10",
	pages = "104028",
	year = "2020"
}

@article{Reitze2019,
	author = {Reitze, David and others},
	title = {Cosmic Explorer: The {U.S.} Contribution to Gravitational-Wave Astronomy beyond {LIGO}},
	journal = {Bull. Am. Astron. Soc.},
	volume = {51},
	pages = {035},
	year = {2019},
	eprint = {1907.04833},
	archivePrefix = {arXiv},
	primaryClass = {astro-ph.IM}
}

@article{Maggiore2020,
	author = {Maggiore, Michele and others},
	title = {Science case for the {Einstein Telescope}},
	journal = {J. Cosmol. Astropart. Phys.},
	volume = {2020},
	number = {03},
	pages = {050},
	year = {2020},
	doi = {10.1088/1475-7516/2020/03/050},
	eprint = {1912.02622},
	archivePrefix = {arXiv},
	primaryClass = {gr-qc}
}

@article{Chatziioannou2020,
	author = {Chatziioannou, Katerina},
	title = {Neutron-star tidal deformability and equation-of-state constraints},
	journal = {Gen. Relativ. Gravit.},
	volume = {52},
	pages = {109},
	year = {2020},
	doi = {10.1007/s10714-020-02754-3},
	eprint = {2006.03168},
	archivePrefix = {arXiv},
	primaryClass = {gr-qc}
}

@article{Huxford2024,
	author = {Huxford, Rachael and Kashyap, Rahul and Borhanian, Ssohrab and Dhani, Arnab and Gupta, Ish and Sathyaprakash, B. S.},
	title = {Accuracy of neutron star radius measurement with the next generation of terrestrial gravitational-wave observatories},
	journal = {Phys. Rev. D},
	volume = {109},
	pages = {103035},
	year = {2024},
	doi = {10.1103/PhysRevD.109.103035},
	eprint = {2307.05376},
	archivePrefix = {arXiv},
	primaryClass = {gr-qc}
}

@article{Finstad2023,
	author = {Finstad, Daniel and White, Laurel V. and Brown, Duncan A.},
	title = {Prospects for a Precise Equation of State Measurement from Advanced LIGO and Cosmic Explorer},
	journal = {Astrophys. J.},
	volume = {955},
	pages = {45},
	year = {2023},
	doi = {10.3847/1538-4357/acf12f},
	eprint = {2211.01396},
	archivePrefix = {arXiv},
	primaryClass = {astro-ph.HE}
}

@article{Williams2022,
	author = {Williams, Natalie and Pratten, Geraint and Schmidt, Patricia},
	title = {Prospects for distinguishing dynamical tides in inspiralling binary neutron stars with third generation gravitational-wave detectors},
	journal = {Phys. Rev. D},
	volume = {105},
	pages = {123032},
	year = {2022},
	doi = {10.1103/PhysRevD.105.123032},
	eprint = {2203.00623},
	archivePrefix = {arXiv},
	primaryClass = {gr-qc}
}

@article{Gamba2021,
	author = {Gamba, Rossella and Breschi, Matteo and Bernuzzi, Sebastiano and Agathos, Michalis and Nagar, Alessandro},
	title = {Waveform systematics in the gravitational-wave inference of tidal parameters and equation of state from binary neutron-star signals},
	journal = {Phys. Rev. D},
	volume = {103},
	pages = {124015},
	year = {2021},
	doi = {10.1103/PhysRevD.103.124015},
	eprint = {2009.08467},
	archivePrefix = {arXiv},
	primaryClass = {gr-qc}
}

@article{Steinhoff2021,
	author = {Steinhoff, Jan and Hinderer, Tanja and Dietrich, Tim and Foucart, Fran\c{c}ois},
	title = {Spin effects on neutron star fundamental-mode dynamical tides: Phenomenology and comparison to numerical simulations},
	journal = {Phys. Rev. Research},
	volume = {3},
	pages = {033129},
	year = {2021},
	doi = {10.1103/PhysRevResearch.3.033129},
	eprint = {2103.06100},
	archivePrefix = {arXiv},
	primaryClass = {gr-qc}
}

@article{PitrePoisson2025,
	author = {Pitre, Tristan and Poisson, Eric},
	title = {Impact of nonlinearities on relativistic dynamical tides in compact binary inspirals},
	journal = {Phys. Rev. D},
	volume = {112},
	pages = {084017},
	year = {2025},
	doi = {10.1103/1wwz-6jyc},
	eprint = {2506.08722},
	archivePrefix = {arXiv},
	primaryClass = {gr-qc}
}

@article{BretzYu2026,
	author = {Bretz, Joseph and Yu, Hang},
	title = {Impact on inferred neutron star equation of state due to nonlinear hydrodynamics, background spin, and relativity},
	journal = {Class. Quantum Grav.},
	volume = {43},
	number = {12},
	pages = {125011},
	year = {2026},
	doi = {10.1088/1361-6382/ae76b5},
	eprint = {2602.04951},
	archivePrefix = {arXiv},
	primaryClass = {gr-qc}
}

@article{Walker2024,
	author = {Walker, Kris and Smith, Rory and Thrane, Eric and Reardon, Daniel J.},
	title = {Precision constraints on the neutron star equation of state with third-generation gravitational-wave observatories},
	journal = {Phys. Rev. D},
	volume = {110},
	pages = {043013},
	year = {2024},
	doi = {10.1103/PhysRevD.110.043013}
}

@article{Abbott2018,
	author = {{LIGO Scientific Collaboration} and {Virgo Collaboration}},
	title = {GW170817: Measurements of Neutron Star Radii and Equation of State},
	journal = {Phys. Rev. Lett.},
	volume = {121},
	pages = {161101},
	year = {2018},
	doi = {10.1103/PhysRevLett.121.161101}
}

@article{FlanaganHinderer2008,
	author = {Flanagan, Eanna E. and Hinderer, Tanja},
	title = {Constraining neutron-star tidal Love numbers with gravitational-wave detectors},
	journal = {Phys. Rev. D},
	volume = {77},
	pages = {021502(R)},
	year = {2008},
	doi = {10.1103/PhysRevD.77.021502}
}

@article{Hinderer2008,
	author = {Hinderer, Tanja},
	title = {Tidal Love numbers of neutron stars},
	journal = {Astrophys. J.},
	volume = {677},
	pages = {1216--1220},
	year = {2008},
	doi = {10.1086/533487}
}

@article{Lai1994,
	author = {Lai, Dong},
	title = {Resonant oscillations and tidal heating in coalescing binary neutron stars},
	journal = {Mon. Not. R. Astron. Soc.},
	volume = {270},
	pages = {611--629},
	year = {1994},
	doi = {10.1093/mnras/270.3.611}
}

@article{Hinderer2016,
	author = {Hinderer, Tanja and others},
	title = {Effects of Neutron-Star Dynamic Tides on Gravitational Waveforms within the Effective-One-Body Approach},
	journal = {Phys. Rev. Lett.},
	volume = {116},
	pages = {181101},
	year = {2016},
	doi = {10.1103/PhysRevLett.116.181101}
}

@article{Hinderer2010,
	author = {Hinderer, Tanja and Lackey, Benjamin D. and Lang, Ryan N. and Read, Jocelyn S.},
	title = {Tidal deformability of neutron stars with realistic equations of state and their gravitational wave signatures in binary inspiral},
	journal = {Phys. Rev. D},
	volume = {81},
	pages = {123016},
	year = {2010},
	doi = {10.1103/PhysRevD.81.123016}
}

@article{Steinhoff2016,
	author = {Steinhoff, Jan and Hinderer, Tanja and Buonanno, Alessandra and Taracchini, Andrea},
	title = {Dynamical tides in general relativity: Effective action and effective-one-body Hamiltonian},
	journal = {Phys. Rev. D},
	volume = {94},
	pages = {104028},
	year = {2016},
	doi = {10.1103/PhysRevD.94.104028}
}

@article{SchmidtHinderer2019,
	author = {Schmidt, Patricia and Hinderer, Tanja},
	title = {Frequency domain model of f-mode dynamic tides in gravitational waveforms from compact binary inspirals},
	journal = {Phys. Rev. D},
	volume = {100},
	pages = {021501(R)},
	year = {2019},
	doi = {10.1103/PhysRevD.100.021501}
}

@article{Hegade2024,
	author = {Hegade K. R., Abhishek and Ripley, Justin L. and Yunes, Nicol\'as},
	title = {Dynamical tidal response of nonrotating relativistic stars},
	journal = {Phys. Rev. D},
	volume = {109},
	pages = {104064},
	year = {2024},
	doi = {10.1103/PhysRevD.109.104064}
}

@article{YuSpin2024,
	author = {Yu, Hang and Arras, Phil and Weinberg, Nevin N.},
	title = {Dynamical tides during the inspiral of rapidly spinning neutron stars: Solutions beyond mode resonance},
	journal = {Phys. Rev. D},
	volume = {110},
	pages = {024039},
	year = {2024},
	doi = {10.1103/PhysRevD.110.024039}
}

@article{Kuan2025,
	author = {Kuan, Hao-Jui and Kiuchi, Kenta and Shibata, Masaru},
	title = {Tidal Resonance in Binary Neutron Star Inspirals: A High-Precision Study in Numerical Relativity},
	journal = {Phys. Rev. Lett.},
	volume = {135},
	pages = {141403},
	year = {2025},
	doi = {10.1103/j3zk-z17h}
}

@article{Schenk2002,
	author = {Schenk, A. Katrin and Arras, Phil and Flanagan, Eanna E. and Teukolsky, Saul A. and Wasserman, Ira},
	title = {Nonlinear mode coupling in rotating stars and the r-mode instability in neutron stars},
	journal = {Phys. Rev. D},
	volume = {65},
	pages = {024001},
	year = {2002},
	doi = {10.1103/PhysRevD.65.024001}
}

@article{Weinberg2012,
	author = {Weinberg, Nevin N. and Arras, Phil and Quataert, Eliot and Burkart, Joshua},
	title = {Nonlinear Tides in Close Binary Systems},
	journal = {Astrophys. J.},
	volume = {751},
	pages = {136},
	year = {2012},
	doi = {10.1088/0004-637X/751/2/136}
}

@article{Yu2023,
	author = {Yu, Hang and Weinberg, Nevin N. and Arras, Phil and Kwon, James and Venumadhav, Tejaswi},
	title = {Beyond the linear tide: impact of the non-linear tidal response of neutron stars on gravitational waveforms from binary inspirals},
	journal = {Mon. Not. R. Astron. Soc.},
	volume = {519},
	number = {3},
	pages = {4325--4343},
	year = {2023},
	doi = {10.1093/mnras/stac3614}
}

@misc{Yu2026,
	author = {Yu, Hang and Nicolini, Giorgio and Lau, Shu Yan and Kwon, K. J. and Venumadhav, Tejaswi and Andersson, Nils and Pnigouras, Pantelis and Gittins, Fabian and Nanda, Amlan},
	title = {Nonlinear hydrodynamics in spinning neutron stars: Theoretical universal relations and equilibrium solutions},
	year = {2026},
	eprint = {2607.07943},
	archivePrefix = {arXiv},
	primaryClass = {gr-qc}
}

@article{Stone2020,
	author = {Stone, James M. and Tomida, Kengo and White, Christopher J. and Felker, Kyle G.},
	title = {The Athena++ Adaptive Mesh Refinement Framework: Design and Magnetohydrodynamic Solvers},
	journal = {Astrophys. J. Suppl. Ser.},
	volume = {249},
	pages = {4},
	year = {2020},
	doi = {10.3847/1538-4365/ab929b}
}

@article{Townsend2013,
	author = {Townsend, R. H. D. and Teitler, S. A.},
	title = {GYRE: an open-source stellar oscillation code based on a new Magnus Multiple Shooting scheme},
	journal = {Mon. Not. R. Astron. Soc.},
	volume = {435},
	pages = {3406},
	year = {2013},
	doi = {10.1093/mnras/stt1533}
}

@article{AnderssonPnigouras2021,
	author = {Andersson, Nils and Pnigouras, Pantelis},
	title = {The phenomenology of dynamical neutron star tides},
	journal = {Mon. Not. R. Astron. Soc.},
	volume = {503},
	number = {1},
	pages = {533--539},
	year = {2021},
	doi = {10.1093/mnras/stab371}
}

@article{Bernuzzi2015,
	author = {Bernuzzi, Sebastiano and Nagar, Alessandro and Dietrich, Tim and Damour, Thibault},
	title = {Modeling the Dynamics of Tidally Interacting Binary Neutron Stars up to the Merger},
	journal = {Phys. Rev. Lett.},
	volume = {114},
	number = {16},
	pages = {161103},
	year = {2015},
	doi = {10.1103/PhysRevLett.114.161103}
}

@article{BinningtonPoisson2009,
	author = {Binnington, Taylor and Poisson, Eric},
	title = {Relativistic theory of tidal Love numbers},
	journal = {Phys. Rev. D},
	volume = {80},
	number = {8},
	pages = {084018},
	year = {2009},
	doi = {10.1103/PhysRevD.80.084018}
}

@article{DamourNagar2009,
	author = {Damour, Thibault and Nagar, Alessandro},
	title = {Relativistic tidal properties of neutron stars},
	journal = {Phys. Rev. D},
	volume = {80},
	number = {8},
	pages = {084035},
	year = {2009},
	doi = {10.1103/PhysRevD.80.084035}
}

@article{Dietrich2017,
	author = {Dietrich, Tim and Bernuzzi, Sebastiano and Tichy, Wolfgang},
	title = {Closed-form tidal approximants for binary neutron star gravitational waveforms constructed from high-resolution numerical relativity simulations},
	journal = {Phys. Rev. D},
	volume = {96},
	number = {12},
	pages = {121501},
	year = {2017},
	doi = {10.1103/PhysRevD.96.121501}
}

@article{Dietrich2019,
	author = {Dietrich, Tim and Samajdar, Anuradha and Khan, Sebastian and Johnson-McDaniel, Nathan K. and Dudi, Reetika and Tichy, Wolfgang},
	title = {Improving the {NRTidal} model for binary neutron star systems},
	journal = {Phys. Rev. D},
	volume = {100},
	number = {4},
	pages = {044003},
	year = {2019},
	doi = {10.1103/PhysRevD.100.044003}
}

@article{Pan2020,
	author = {Pan, Zhen and Lyu, Zhenwei and Bonga, B\'eatrice and Ortiz, N\'estor and Yang, Huan},
	title = {Probing Crust Meltdown in Inspiraling Binary Neutron Stars},
	journal = {Phys. Rev. Lett.},
	volume = {125},
	number = {20},
	pages = {201102},
	year = {2020},
	doi = {10.1103/PhysRevLett.125.201102}
}

@article{Pratten2020,
	author = {Pratten, Geraint and Schmidt, Patricia and Hinderer, Tanja},
	title = {Gravitational-wave asteroseismology with fundamental modes from compact binary inspirals},
	journal = {Nat. Commun.},
	volume = {11},
	number = {1},
	pages = {2553},
	year = {2020},
	doi = {10.1038/s41467-020-15984-5}
}

@article{Pratten2022,
	author = {Pratten, Geraint and Schmidt, Patricia and Williams, Natalie},
	title = {Impact of Dynamical Tides on the Reconstruction of the Neutron Star Equation of State},
	journal = {Phys. Rev. Lett.},
	volume = {129},
	number = {8},
	pages = {081102},
	year = {2022},
	doi = {10.1103/PhysRevLett.129.081102}
}

@article{Vines2011,
	author = {Vines, Justin and Flanagan, {\'E}anna {\'E}. and Hinderer, Tanja},
	title = {Post-1-{Newtonian} tidal effects in the gravitational waveform from binary inspirals},
	journal = {Phys. Rev. D},
	volume = {83},
	number = {8},
	pages = {084051},
	year = {2011},
	doi = {10.1103/PhysRevD.83.084051}
}

@article{HoLai1999,
	author = {Ho, Wynn C. G. and Lai, Dong},
	title = {Resonant tidal excitations of rotating neutron stars in coalescing binaries},
	journal = {Mon. Not. R. Astron. Soc.},
	volume = {308},
	number = {1},
	pages = {153--166},
	year = {1999},
	doi = {10.1046/j.1365-8711.1999.02703.x},
	eprint = {astro-ph/9812116},
	archivePrefix = {arXiv},
	url = {https://doi.org/10.1046/j.1365-8711.1999.02703.x}
}

@article{KokkotasSchafer1995,
	author = {Kokkotas, Kostas D. and Sch{\"a}fer, Gerhard},
	title = {Tidal and tidal-resonant effects in coalescing binaries},
	journal = {Mon. Not. R. Astron. Soc.},
	volume = {275},
	number = {2},
	pages = {301--308},
	year = {1995},
	doi = {10.1093/mnras/275.2.301},
	eprint = {gr-qc/9502034},
	archivePrefix = {arXiv},
	url = {https://doi.org/10.1093/mnras/275.2.301}
}

@article{LaiRasioShapiro1994,
	author = {Lai, Dong and Rasio, Frederic A. and Shapiro, Stuart L.},
	title = {Hydrodynamic instability and coalescence of binary neutron stars},
	journal = {Astrophys. J.},
	volume = {420},
	pages = {811--829},
	year = {1994},
	doi = {10.1086/173606},
	eprint = {astro-ph/9304027},
	archivePrefix = {arXiv},
	url = {https://doi.org/10.1086/173606}
}

@article{LaiShapiro1995,
	author = {Lai, Dong and Shapiro, Stuart L.},
	title = {Hydrodynamics of coalescing binary neutron stars: Ellipsoidal treatment},
	journal = {Astrophys. J.},
	volume = {443},
	number = {2},
	pages = {705--716},
	year = {1995},
	doi = {10.1086/175562},
	eprint = {astro-ph/9408054},
	archivePrefix = {arXiv},
	url = {https://doi.org/10.1086/175562}
}

@article{Venumadhav2014,
	author = {Venumadhav, Tejaswi and Zimmerman, Aaron and Hirata, Christopher M.},
	title = {The Stability of Tidally Deformed Neutron Stars to Three- and Four-mode Coupling},
	journal = {Astrophys. J.},
	volume = {781},
	number = {1},
	pages = {23},
	year = {2014},
	doi = {10.1088/0004-637X/781/1/23},
	eprint = {1307.2890},
	archivePrefix = {arXiv},
	primaryClass = {astro-ph.HE},
	url = {https://doi.org/10.1088/0004-637X/781/1/23}
}

@article{Weinberg2013,
	author = {Weinberg, Nevin N. and Arras, Phil and Burkart, Joshua},
	title = {An Instability due to the Nonlinear Coupling of {p}-modes to {g}-modes: Implications for Coalescing Neutron Star Binaries},
	journal = {Astrophys. J.},
	volume = {769},
	number = {2},
	pages = {121},
	year = {2013},
	doi = {10.1088/0004-637X/769/2/121},
	eprint = {1302.2292},
	archivePrefix = {arXiv},
	primaryClass = {astro-ph.SR},
	url = {https://doi.org/10.1088/0004-637X/769/2/121}
}

@article{Weinberg2016,
	author = {Weinberg, Nevin N.},
	title = {Growth Rate of the Tidal {p}-mode {g}-mode Instability in Coalescing Binary Neutron Stars},
	journal = {Astrophys. J.},
	volume = {819},
	number = {2},
	pages = {109},
	year = {2016},
	doi = {10.3847/0004-637X/819/2/109},
	eprint = {1509.06975},
	archivePrefix = {arXiv},
	primaryClass = {astro-ph.SR},
	url = {https://doi.org/10.3847/0004-637X/819/2/109}
}

\clearpage
\onecolumngrid
\raggedbottom
\setcounter{section}{0}
\setcounter{subsection}{0}
\setcounter{equation}{0}
\setcounter{figure}{0}
\renewcommand{\theequation}{S\arabic{equation}}
\renewcommand{\thefigure}{S\arabic{figure}}
\renewcommand{\theHequation}{S\arabic{equation}}
\renewcommand{\theHfigure}{S\arabic{figure}}
\begin{center}
	{\large\bfseries Supplemental Material}
\end{center}
\vspace{-\baselineskip}

\section{Mode coupling model}

The mode coupling model relates the fluid quadrupole spectra to
interactions among the $l=0$ radial mode and the $l=2,4$ f-mode multiplets.
We give the explicit form of Eq.~\eqref{eq:mode} used in
Figs.~\ref{fig:modes}--\ref{fig:frequency}, following
Refs.~\cite{Schenk2002,Weinberg2012,Yu2023,Yu2026}.

We use the displacement coordinates
\begin{equation}
	\begin{aligned}
		X_{lm}&=c_{lm}+c_{l,-m}^*,\qquad
		\widetilde X_{lm}=e^{\ii m\phi}X_{lm},\qquad l=2,4,\\
		x_0&=X_{20},\qquad x_r=c_r+c_r^*,\qquad
		b=\widetilde X_{22},\qquad S=|b|^2,
	\end{aligned}
	\label{eq:s-coordinates}
\end{equation}
where $m=0,\pm2$ for $l=2$ and $m=0,\pm2,\pm4$ for $l=4$.
These coordinates satisfy $X_{l,-m}=X_{lm}^*$; $x_0$ and $x_r$ are real,
while $b$ has the orbital phase removed.
The $l=0,2,4$ canonical mode amplitudes obey the coupled equations in the inertial frame
\begin{align}
	\dot c_{20}&=-\ii\omega_fc_{20}+\ii\omega_f(\mathcal F_0+\mathcal L_{20}),
	\nonumber\\
	\dot c_r&=-\ii\omega_rc_r+\ii\omega_r\mathcal F_r,
	\nonumber\\
	\dot c_{22}&=-\ii\omega_fc_{22}
	+\ii\omega_f(e^{-2\ii\phi}\mathcal F_2+\mathcal L_{22}),
	\nonumber\\
	\dot c_{2,-2}^{*}&=+\ii\omega_fc_{2,-2}^{*}
	-\ii\omega_f(e^{-2\ii\phi}\mathcal F_2+\mathcal L_{22}),
	\nonumber\\
	\dot c_{4m}&=-\ii\omega_4c_{4m}+\ii\omega_4\mathcal L_{4m},
	\nonumber\\
	\dot c_{4,-m}^{*}&=+\ii\omega_4c_{4,-m}^{*}
	-\ii\omega_4\mathcal L_{4m},\qquad m=0,2,4.
	\label{eq:s-mode-system}
\end{align}
The sources $\mathcal F_0$, $\mathcal F_r$, and $\mathcal F_2$
contain the $l=0,2$ interactions, while $\mathcal L_{lm}$ couples the
$l=2$ and $l=4$ modes.  With $a_0=\sqrt{\pi/5}$ and
$a_2=\sqrt{3\pi/10}$, these sources are
\begin{align}
	\mathcal F_0={}&-a_0I_f\epsilon+2\kappa_2S-\kappa_2x_0^2
	+2\kappa_{2r}x_0x_r
	\nonumber\\
	&+a_2J_2(b+b^*)\epsilon+a_0J_2x_0\epsilon-a_0J_{2r}x_r\epsilon
	\nonumber\\
	&+6\zeta_{20}Sx_0+6\zeta_{20r}Sx_r+\zeta_0x_0^3
	-3\zeta_{20r}x_0^2x_r+3\zeta_{0r}x_0x_r^2,
	\label{eq:s-f0}\\
	\mathcal F_r={}&2\kappa_{2r}S+\kappa_{2r}x_0^2+\kappa_{rr}x_r^2
	+a_2J_{2r}(b+b^*)\epsilon-a_0J_{2r}x_0\epsilon
	\nonumber\\
	&+6\zeta_{20r}Sx_0+6\zeta_{2r}Sx_r-\zeta_{20r}x_0^3
	+3\zeta_{0r}x_0^2x_r,
	\label{eq:s-fr}\\
	\mathcal F_2={}&a_2I_f\epsilon
	+(2\kappa_2x_0+2\kappa_{2r}x_r)b-a_0J_2\epsilon b
	\nonumber\\
	&+a_2(J_2x_0+J_{2r}x_r)\epsilon+3\zeta_{22}Sb
	\nonumber\\
	&+(3\zeta_{20}x_0^2+6\zeta_{20r}x_0x_r
	+3\zeta_{2r}x_r^2)b,
	\label{eq:s-f2}\\
	\mathcal L_{lm}={}&
	\frac{\partial\mathcal G_{224}}{\partial X_{l,-m}}
	+e^{-\ii m\phi}
	\frac{\partial\mathcal G_{U,24}}{
		\partial\widetilde X_{l,-m}},\qquad l=2,4,
	\label{eq:s-l4-source}
\end{align}
where $\mathcal G_{224}$ describes coupling between two $l=2$ modes and one
$l=4$ mode, and $\mathcal G_{U,24}$ contains their quadratic tidal overlap,
\begin{align}
	\mathcal G_{224}
	&=\sum_{m_1,m_2=0,\pm2}
	\kappa_{224}^{m_1,m_2,-m_1-m_2}
	X_{2m_1}X_{2m_2}X_{4,-m_1-m_2},
	\label{eq:s-g224}\\
	\mathcal G_{U,24}
	&=\epsilon\sum_{\mu=0,\pm2}W_{2\mu}
	\sum_{m_2,m_4}J_{24}^{m_2,m_4;\mu}
	\widetilde X_{2m_2}\widetilde X_{4m_4}.
	\label{eq:s-gu24}
\end{align}
Here $J_{24}^{m_2,m_4;\mu}=J_{(2,m_2)(4,m_4);2\mu}$, and the three
superscripts of $\kappa_{224}$ are the azimuthal indices of the participating
modes.

In particular, the three- and four-mode coefficients $\kappa_{abc}$ and $\zeta_{abcd}$
are the mode projections of the nonlinear fluid forces~\cite{Weinberg2012,Yu2023,Weinberg2016},
\begin{equation}
	\kappa_{abc}=\frac{1}{E_*}\int_{\mathcal V_*}
	\bm\xi_a^*\!\cdot\bm f_2[\bm\xi_b^*,\bm\xi_c^*]\,d^3x,
	\qquad
	\zeta_{abcd}=\frac{1}{E_*}\int_{\mathcal V_*}
	\bm\xi_a^*\!\cdot\bm f_3[\bm\xi_b^*,\bm\xi_c^*,\bm\xi_d^*]\,d^3x.
	\label{eq:s-coupling-definitions}
\end{equation}
Here $\bm f_2$ and $\bm f_3$ are the quadratic and cubic force densities
in the Lagrangian displacement expansion,
$\bm f_{\rm NL}[\bm\xi]=\bm f_2[\bm\xi,\bm\xi]
+\bm f_3[\bm\xi,\bm\xi,\bm\xi]+\cdots$,
including both pressure and self-gravity.
The overlap and coupling coefficients $I_f$, $J$, $\kappa$, and
$\zeta$ can be obtained from Refs.~\cite{Weinberg2012,Yu2026}.

\section{Perturbative f-mode frequency shift}

At fixed $D$ and $\Omega$, we remove the orbital phase through
$b_{lm}=e^{\ii m\Omega t}c_{lm}$.
We collect the transformed amplitudes and their conjugates in
\begin{equation}
	\bm C=\bigl(b_{22},b_{2,-2},b_{20},b_{00},b_{40},b_{42},b_{4,-2},b_{44},b_{4,-4},
	b_{22}^{*},b_{2,-2}^{*},b_{20}^{*},b_{00}^{*},b_{40}^{*},b_{42}^{*},b_{4,-2}^{*},b_{44}^{*},b_{4,-4}^{*}\bigr)^{\mathsf T}.
	\label{eq:s-canonical-state}
\end{equation}
Here $b_{00}=c_r$. We obtain the stationary tide from the mode equations and expand its
amplitudes in the tidal strength,
\begin{equation}
	\dot{\bm C}=\bm F(\bm C;\epsilon),\qquad
	\bm F(\bar{\bm C};\epsilon)=0,\qquad
	\bar b_{lm}=\epsilon\bar b_{lm}^{(1)}+\epsilon^2\bar b_{lm}^{(2)}+O(\epsilon^3).
	\label{eq:s-equilibrium-expansion}
\end{equation}
At first order,
\begin{equation}
	\bar b_{20}^{(1)}=-a_0I_f,\qquad
	\bar b_{2,\pm2}^{(1)}=\frac{\omega_fa_2I_f}{\omega_f\mp2\Omega},\qquad
	\bar b_{00}^{(1)}=\bar b_{4m}^{(1)}=0,
	\label{eq:s-equilibrium-first}
\end{equation}
The second-order equilibrium coefficients required for the frequency shift are
\begin{equation}
	\begin{aligned}
		\bar b_{20}^{(2)}={}&2\kappa_2\bigl(\bar b_{22}^{(1)}+\bar b_{2,-2}^{(1)}\bigr)^2-4\kappa_2(\bar b_{20}^{(1)})^2
		+2a_2J_2\bigl(\bar b_{22}^{(1)}+\bar b_{2,-2}^{(1)}\bigr)+2a_0J_2\bar b_{20}^{(1)},\\
		\bar b_{00}^{(2)}={}&2\kappa_{2r}\bigl(\bar b_{22}^{(1)}+\bar b_{2,-2}^{(1)}\bigr)^2+4\kappa_{2r}(\bar b_{20}^{(1)})^2
		+2a_2J_{2r}\bigl(\bar b_{22}^{(1)}+\bar b_{2,-2}^{(1)}\bigr)-2a_0J_{2r}\bar b_{20}^{(1)},\\
		\bar b_{40}^{(2)}={}&2\kappa_{224}^{2,-2,0}\bigl(\bar b_{22}^{(1)}+\bar b_{2,-2}^{(1)}\bigr)^2+4\kappa_{224}^{0,0,0}(\bar b_{20}^{(1)})^2
		+2a_2J_{24}^{2,0;-2}\bigl(\bar b_{22}^{(1)}+\bar b_{2,-2}^{(1)}\bigr)-2a_0J_{24}^{0,0;0}\bar b_{20}^{(1)} .
	\end{aligned}
	\label{eq:s-equilibrium-second}
\end{equation}

Small oscillations about this equilibrium satisfy
$\delta\dot{\bm C}=\mathsf J(\epsilon)\delta\bm C$, where
$\mathsf J=\left.\partial\bm F/\partial\bm C\right|_{\bar{\bm C}}$.
For a normal mode component $\delta\bm C=\bm v_0e^{-\ii\sigma t}$,
\begin{equation}
	\mathsf J(\epsilon)\bm v_0=-\ii\sigma\bm v_0,\qquad
	2\pi f_{22}=\sigma+2\Omega,\qquad
	\sigma^{(0)}=\omega_f-2\Omega.
	\label{eq:s-canonical-eigenproblem}
\end{equation}
The branch connected to the $m=2$ f-mode of the isolated star has
\begin{equation}
	\bm v_0^{(0)}=(1,0,\ldots,0)^{\mathsf T},\qquad
	\bm l_0^{(0)\dagger}=(1,0,\ldots,0).
	\label{eq:s-zeroth-order-eigenvectors}
\end{equation}
We expand the equilibrium Jacobian, eigenfrequency, and eigenvector as
\begin{equation}
	\begin{aligned}
		\mathsf J&=\mathsf J^{(0)}+\epsilon\mathsf J^{(1)}+\epsilon^2\mathsf J^{(2)}+O(\epsilon^3),\\
		\sigma&=\sigma^{(0)}+\epsilon\sigma^{(1)}+\epsilon^2\sigma^{(2)}+O(\epsilon^3),\\
		\bm v_0&=\bm v_0^{(0)}+\epsilon\bm v_0^{(1)}
		+\epsilon^2\bm v_0^{(2)}+O(\epsilon^3).
	\end{aligned}
	\label{eq:s-jacobian-expansion}
\end{equation}
We fix the eigenvector normalization by
$\bm l_0^{(0)\dagger}\bm v_0^{(n)}=0$ for $n\geq1$.
Matching the first two orders in Eq.~\eqref{eq:s-canonical-eigenproblem} gives
\begin{equation}
	\begin{aligned}
		\mathsf J^{(0)}\bm v_0^{(1)}+\mathsf J^{(1)}\bm v_0^{(0)}
		&=-\ii\sigma^{(0)}\bm v_0^{(1)}-\ii\sigma^{(1)}\bm v_0^{(0)},\\
		\mathsf J^{(0)}\bm v_0^{(2)}+\mathsf J^{(1)}\bm v_0^{(1)}+\mathsf J^{(2)}\bm v_0^{(0)}
		&=-\ii\sigma^{(0)}\bm v_0^{(2)}-\ii\sigma^{(1)}\bm v_0^{(1)}
		-\ii\sigma^{(2)}\bm v_0^{(0)}.
	\end{aligned}
	\label{eq:s-order-equations}
\end{equation}
Left multiplication by $\bm l_0^{(0)\dagger}$ cancels the terms containing
$\mathsf J^{(0)}$ against those proportional to $\sigma^{(0)}$.
Using the normalization above then gives Eq.~\eqref{eq:eigenvector-frequency-shifts} in the main text,
\begin{equation}
	\begin{aligned}
		\frac{2\pi\Delta f^{(1)}}{\epsilon}=\sigma^{(1)}
		&=\ii\bm l_0^{(0)\dagger}\mathsf J^{(1)}\bm v_0^{(0)},\\
		\frac{2\pi\Delta f^{(2)}}{\epsilon^2}=\sigma^{(2)}
		&=\ii\bm l_0^{(0)\dagger}\mathsf J^{(2)}\bm v_0^{(0)}
		+\ii\bm l_0^{(0)\dagger}\mathsf J^{(1)}\bm v_0^{(1)}.
	\end{aligned}
	\label{eq:s-projected-frequency-shifts}
\end{equation}

For each remaining eigenstate, $\sigma_\nu^{(0)}$ denotes the zeroth-order
eigenfrequency, and we choose
$\bm l_\mu^{(0)\dagger}\bm v_\nu^{(0)}=\delta_{\mu\nu}$.
Projecting the first line of Eq.~\eqref{eq:s-order-equations} onto
$\bm l_\nu^{(0)\dagger}$ gives
\begin{equation}
	\bm v_0^{(1)}=\ii\sum_{\nu\ne0}
	\frac{\bm l_\nu^{(0)\dagger}\mathsf J^{(1)}\bm v_0^{(0)}}
	{\sigma^{(0)}-\sigma_\nu^{(0)}}\,\bm v_\nu^{(0)}.
	\label{eq:s-first-order-eigenvector}
\end{equation}
Evaluating these projections in the ordering of Eq.~\eqref{eq:s-canonical-state} yields
\begin{equation}
	\begin{aligned}
		\bm v_0^{(1)}=\Bigl(&0,0,
		\frac{\omega_fg_0}{\omega_f-\sigma^{(0)}},
		\frac{\omega_rg_r}{\omega_r-\sigma^{(0)}},
		\frac{\omega_4G_0}{\omega_4-\sigma^{(0)}},
		\frac{\omega_4G_2}{\omega_4-\sigma^{(0)}-2\Omega},0,
		\frac{\omega_4G_4}{\omega_4-\sigma^{(0)}-4\Omega},0,\\[5pt]
		&0,-\frac{a_0(J_2+4\kappa_2I_f)}{2},
		\frac{\omega_fg_0}{\omega_f+\sigma^{(0)}},
		\frac{\omega_rg_r}{\omega_r+\sigma^{(0)}},
		\frac{\omega_4G_0}{\omega_4+\sigma^{(0)}},0,
		\frac{\omega_4G_2}{\omega_4+\sigma^{(0)}+2\Omega},0,
		\frac{\omega_4G_4}{\omega_4+\sigma^{(0)}+4\Omega}\Bigr)^{\mathsf T},
	\end{aligned}
	\label{eq:s-explicit-first-order-eigenvector}
\end{equation}
where
\begin{equation}
	\begin{aligned}
		g_0={}&a_2J_2+2\kappa_2\bigl(\bar b_{22}^{(1)}+\bar b_{2,-2}^{(1)}\bigr),\qquad
		g_r=a_2J_{2r}+2\kappa_{2r}\bigl(\bar b_{22}^{(1)}+\bar b_{2,-2}^{(1)}\bigr),\\
		G_0={}&a_2J_{24}^{2,0;-2}+2\kappa_{224}^{2,-2,0}\bigl(\bar b_{22}^{(1)}+\bar b_{2,-2}^{(1)}\bigr),\qquad
		G_2=-a_0J_{24}^{2,-2;0}+4\kappa_{224}^{0,2,-2}\bar b_{20}^{(1)},\qquad
		G_4=a_2J_{24}^{2,-4;2}+2\kappa_{224}^{2,2,-4}\bigl(\bar b_{22}^{(1)}+\bar b_{2,-2}^{(1)}\bigr).
	\end{aligned}
	\label{eq:s-mixing}
\end{equation}
Expanding the equilibrium Jacobian using Eqs.~\eqref{eq:s-equilibrium-first}
and \eqref{eq:s-equilibrium-second} determines $\mathsf J^{(1)}$ and $\mathsf J^{(2)}$.
Substituting these matrices and the eigenvectors in Eqs.~\eqref{eq:s-zeroth-order-eigenvectors}
and \eqref{eq:s-explicit-first-order-eigenvector} into Eq.~\eqref{eq:s-projected-frequency-shifts}
gives the explicit frequency shifts,
\begin{equation}
	\begin{aligned}
		\frac{\Delta f^{(1)}}{\epsilon f_{\rm lin}}
		={}&a_0(J_2+4\kappa_2I_f),\\
		\frac{\Delta f^{(2)}}{\epsilon^2 f_{\rm lin}}
		={}&-H_{22}-\frac12[a_0(J_2+4\kappa_2I_f)]^2
		-\frac{2\omega_f^2g_0^2}{\omega_f^2-(\sigma^{(0)})^2}
		-\frac{2\omega_r^2g_r^2}{\omega_r^2-(\sigma^{(0)})^2}
		-\sum_{m=0,2,4}\frac{2\omega_4^2G_m^2}{\omega_4^2-(\sigma^{(0)}+m\Omega)^2},
	\end{aligned}
	\label{eq:s-frequency-corrections}
\end{equation}
with $f_{\rm lin}=\omega_f/(2\pi)$,
$f_{22}=f_{\rm lin}+\Delta f^{(1)}+\Delta f^{(2)}+O(\epsilon^3)$, and
\begin{equation}
	H_{22}=4\kappa_2\bar b_{20}^{(2)}+4\kappa_{2r}\bar b_{00}^{(2)}
	+6\zeta_{22}\bigl(\bar b_{22}^{(1)}+\bar b_{2,-2}^{(1)}\bigr)^2+12\zeta_{20}(\bar b_{20}^{(1)})^2
	+4\kappa_{224}^{-2,2,0}\bar b_{40}^{(2)}.
	\label{eq:s-H22}
\end{equation}

\section{Nonlinear tides induced GW phase shift}

The phase calculation in Fig.~\ref{fig:phase} couples the evolving stellar
quadrupole to the orbital motion.  Earlier Newtonian treatments described
this coupling with ellipsoidal stellar models~\cite{LaiRasioShapiro1994,LaiShapiro1995}
and coupled orbit--mode equations~\cite{KokkotasSchafer1995}.
Here the stellar response is supplied by fully nonlinear hydrodynamics or
the mode coupling model, with the orbital radiation reaction of
Ref.~\cite{Yu2023}.

The total mass quadrupole contains orbital and stellar contributions,
\begin{equation}
	\mathcal I_{ij}^{\rm tot}=\mathcal I_{ij}^{\rm orb}+N_*\mathcal I_{ij},
	\qquad
	\mathcal I_{ij}^{\rm orb}=\mu D^2
	\left(n_i n_j-\frac{1}{3}\delta_{ij}\right),
	\label{eq:s-total-quadrupole}
\end{equation}
where $\bm n=(\cos\phi,\sin\phi,0)$ and $N_*=2$ accounts for the
identical tidal deformations of the two neutron stars.
The stellar quadrupole follows directly from the density distribution.
Its symmetric trace-free Cartesian form,
\begin{equation}
	\mathcal I_{ij}(t)=\int_{\mathcal V_*}\rho({\bm x},t)
	\left(x_i x_j-\frac{1}{3}r^2\delta_{ij}\right)d^3x,
	\label{eq:s-stellar-quadrupole}
\end{equation}
is related to the spherical components by
\begin{equation}
	Q_{20}=3\sqrt{\frac{5}{16\pi}}\mathcal I_{zz},\qquad
	Q_{22}=\sqrt{\frac{15}{32\pi}}
	(\mathcal I_{xx}-\mathcal I_{yy}-2\ii\mathcal I_{xy}).
	\label{eq:s-quadrupole-conversion}
\end{equation}
Here, the stellar quadrupole is calculated in the center-of-mass frame of each star.

For the waveform in Fig.~\ref{fig:phase}(a), the total spherical quadrupole is
\begin{equation}
	Q_{22}^{\rm tot}=\sqrt{\frac{15}{32\pi}}\mu D^2e^{-2\ii\phi}+N_*Q_{22}.
	\label{eq:s-waveform-quadrupole}
\end{equation}
The leading mass-quadrupole formula for a face-on observer at distance $d$ is
\addtolength{\abovedisplayskip}{-1pt}
\addtolength{\belowdisplayskip}{-1pt}
\begin{equation}
	h_+-\ii h_\times=-\frac{G}{c^4d}\sqrt{\frac{32\pi}{15}}
	\frac{d^2Q_{22}^{\rm tot}}{dt^2},
	\label{eq:s-quadrupole-waveform}
\end{equation}
where the overall sign fixes a common polarization convention with
$h_+\propto\cos(2\phi)$ for a circular orbit.
We set the initial orbital phases to zero and use $d=100\,\mathrm{Mpc}$.

At leading order, the radiation-reaction acceleration takes the
Burke--Thorne form~\cite{Yu2023},
\begin{equation}
	a_i^{\rm BT}(\bm X)=-\frac{2G}{5c^5}
	\frac{d^5\mathcal I_{ij}^{\rm tot}}{dt^5}X_j,
	\label{eq:s-burke-thorne-field}
\end{equation}
where $\bm X$ denotes position relative to the binary center of mass.
The difference between the accelerations of the stellar centers gives the
radiation reaction of the relative orbit,
\begin{equation}
	a_i^{\rm GW}=-\frac{2G}{5c^5}
	\frac{d^5\mathcal I_{ij}^{\rm tot}}{dt^5}D_j,\qquad
	a_D^{\rm GW}=n_i a_i^{\rm GW},\qquad
	a_\phi^{\rm GW}=\lambda_i a_i^{\rm GW},
	\label{eq:s-relative-radiation-reaction}
\end{equation}
with $\bm\lambda=(-\sin\phi,\cos\phi,0)$.

The trajectory of the companion star relative to the primary star is governed by
\begin{align}
	\ddot D&=D\Omega^2-\frac{G(M_*+M_c)}{D^2}
	+a_D^{\rm tide}+a_D^{\rm GW},\nonumber\\
	\dot\phi&=\Omega,\qquad
	\dot\Omega=\frac{a_\phi^{\rm tide}+a_\phi^{\rm GW}}{D}
	-\frac{2\dot D\Omega}{D},
	\label{eq:s-orbit-evolution}
\end{align}
where $a_D^{\rm tide}$ and $a_\phi^{\rm tide}$ denote the accelerations
induced by tidal deformation, as given in
Eq.~\eqref{eq:tidal-accelerations} of the main text.
We decompose $\bm a^{\rm GW}$ into a point-particle contribution and
tidal corrections.
Including the leading tidal corrections gives~\cite{Yu2023}
\begin{align}
	a_D^{\rm GW}&=a_D^{\rm pp},\nonumber\\
	a_\phi^{\rm GW}&=a_\phi^{\rm pp}
	-\frac{128N_*G}{5c^5}\sqrt{\frac{2\pi}{15}}D\Omega^5
	\operatorname{Re}\left(e^{2\ii\phi}Q_{22}\right)-\frac{96N_*G^2M_c\Omega^3}{5c^5D^2}
	\left[W_{22}\operatorname{Re}\left(e^{2\ii\phi}Q_{22}\right)
	+\frac{W_{20}}{2}Q_{20}\right],
	\label{eq:s-tidal-radiation-reaction}
\end{align}
where the point-particle contribution is
\begin{align}
	a_D^{\rm pp}&=\frac{16G^2M_*M_c}{5c^5D^3}\dot D
	\left[\dot D^2+6D^2\Omega^2+\frac{4G(M_*+M_c)}{3D}\right],\nonumber\\
	a_\phi^{\rm pp}&=\frac{8G^2M_*M_c}{5c^5D^2}\Omega
	\left[9\dot D^2-6D^2\Omega^2+\frac{2G(M_*+M_c)}{D}\right].
	\label{eq:s-point-particle-reaction}
\end{align}
The two tidal corrections in Eq.~\eqref{eq:s-tidal-radiation-reaction}
account for radiation from the
stellar quadrupole and the tidal modification of the orbital radiation,
respectively.  They are evaluated in the slowly evolving, nearly circular
orbit approximation.
Equations~\eqref{eq:s-orbit-evolution}--\eqref{eq:s-point-particle-reaction}
couple the orbital motion to $Q_{20}(t)$ and $Q_{22}(t)$, which are
determined by the stellar evolution below.

With the radiation-reaction acceleration $\bm a^{\rm RR}$, the momentum
and energy equations in Eq.~\eqref{eq:hydrodynamics} take the form
\begin{align}
	\partial_t(\rho\bm v)+\bm\nabla\!\cdot
	(\rho\bm v\bm v+P\bm I)
	&=-\rho\bm\nabla(\Phi+U_{\rm tide})+\rho\bm a^{\rm RR},\nonumber\\
	\partial_tE+\bm\nabla\!\cdot[(E+P)\bm v]
	&=-\rho\bm v\!\cdot\bm\nabla(\Phi+U_{\rm tide})
	+\rho\bm v\!\cdot\bm a^{\rm RR}.
	\label{eq:s-hydro-inspiral}
\end{align}
The continuity equation, Poisson equation, and equation of state retain
their forms in the main text.  The tidal potential in Eq.~\eqref{eq:tide}
is evaluated along the evolving orbit $D(t),\phi(t)$.
In the stellar frame, subtracting the acceleration of the stellar center
$\bm X_*$ gives
\begin{equation}
	a_i^{\rm RR}(\bm x)
	=a_i^{\rm BT}(\bm X_*+\bm x)-a_i^{\rm BT}(\bm X_*)
	=a_i^{\rm orb}+N_*a_i^{\rm self},
	\label{eq:s-orbital-reaction-fluid}
\end{equation}
where
\begin{equation}
	a_i^{\rm orb}=-\frac{2G}{5c^5}
	\frac{d^5\mathcal I_{ij}^{\rm orb}}{dt^5}x_j,\qquad
	a_i^{\rm self}=-\frac{2G}{5c^5}
	\frac{d^5\mathcal I_{ij}}{dt^5}x_j.
	\label{eq:s-radiation-reaction-components}
\end{equation}

\end{document}